\documentclass[12pt,a4paper]{iopart}

\expandafter\let\csname equation*\endcsname\relax
\expandafter\let\csname endequation*\endcsname\relax
\usepackage{hyperref}

\usepackage{amsmath}    
\usepackage{amssymb}    
\usepackage{color}      
\usepackage{graphicx}   
\usepackage{bm}
\usepackage[sort&compress,numbers]{natbib}

\begin{document}

\title{Is the Kinetoplast DNA a Percolating Network of Linked Rings at its Critical Point?}

\author{%
Davide Michieletto\,$^{1,*}$,
Davide Marenduzzo\,$^{2}$
and Enzo Orlandini\,$^{3}$%
\footnote{To whom correspondence should be addressed.
Email: d.michieletto@warwick.ac.uk}}

\address{%
$^{1}$ Department of Physics and Centre for Complexity Science, University of Warwick, Coventry CV4 7AL, United Kingdom.
$^{2}$ School of Physics and Astronomy, University of Edinburgh, Mayfield Road, Edinburgh EH9 3JZ, Scotland, United Kingdom.
$^{3}$ Dipartimento di Fisica e Astronomia, Sezione INFN, Universit\`a di Padova, Via Marzolo 8, 35131 Padova, Italy.}

\begin{abstract}
In this work we present a computational study of the Kinetoplast
genome, modelled as a large number of semiflexible unknotted loops, which 
are allowed to link with each other. As the DNA density increases, the systems shows a percolation transition between a gas of unlinked rings and a network of linked loops which spans the whole system. Close to the percolation transition, we find that the mean valency of the network, i.e. the average number of loops which are linked to any one loop, is around 3 as found experimentally for the Kinetoplast DNA. 
Even more importantly, by simulating the digestion of the network by a 
restriction enzyme, we show that the distribution of oligomers, i.e. 
structures formed by a few loops which remain linked after digestion, 
quantitatively matches experimental data obtained from
gel electrophoresis, provided that the density is, once again, close
to the percolation transition. With respect to previous
work, our analysis builds on a reduced number of assumptions, yet can
still fully explain the experimental data.
Our findings suggest that the Kinetoplast DNA can be viewed as a network of linked loops positioned very close to the percolation transition, and we discuss the possible biological implications of this remarkable fact.
\end{abstract}

\section{Introduction}

A Kinetoplast~\cite{Fairlamb1978} is a network of linked DNA loops commonly found in a group of unicellular eukaryotic organisms of the class Kinetoplastida. Some of these organisms are responsible for important diseases such as sleeping sickness and leishmaniasis~\cite{Young1987,Jacobson2003,MacLean2004}. The Kinetoplast DNA (kDNA) is known for its unique structure. Thousands of short ($1-2.5$ $kbp$) DNA loops are interlinked, forming a spanning network that fills the mitochondria. The short loops, or mini-circles, are also linked with few large circles, or maxi-circles, consisting of around $30-50$ $kbp$ \cite{Shapiro1995}. The loops are found to be in a relaxed state, \emph{i.e.} they are not supercoiled, contrarily to DNA loops in other similar organisms.  \emph{C. fasciculata} mini-circles assemble in a network whose shape resembles that of a disk which measures $1$ $\mu m$ in diameter and is $0.4$ $\mu m$ thick~\cite{Perez-Morga1993,Jensen2012}. Networks which are removed from the mitochondria, \emph{e.g.} via cell lysis, expand into an elliptical shape whose minor and major axis are respectively around $10$ $\mu m$ and $15$ $\mu m$, \emph{i.e.} roughly a hundred times bigger than their dimension \emph{in vivo}~\cite{Jensen2012}. This suggests that the networks experience a confinement within the mitochondria. 

It has been observed~\cite{Ogbadoyi2003} that a ``tripartite attachment complex'' (TAC) keeps the Kinetoplast statically in place near the basal body, from which it is physically separated by the mitochondrial envelope. 
Transmission electron microscopy images of Kinetoplast networks \emph{in vivo}~\cite{Lukes2002,Ogbadoyi2003,Gluenz2007a,Lai2008,Docampo2010}  also suggest that the shape of the mitochondrial membrane near the Kinetoplast acts as a physical constraint on the outer structure of the network, while it is likely that histone-like proteins, such as p16, p17 and p18, or ``KAP proteins'' encoded in genes KAP2, KAP3 and KAP4, act as chemical constraint on the inner structure~\cite{Xu1993,Xu1996,Hines1998,Avliyakulov2004}. 

The concentration of DNA in the Kinetoplast has been found to be around $50$  $mg/ml$~\cite{Shapiro1995}, similar to that found in bacteria ($20$ $mg/ml$) but far smaller than the one inside the head of a T4 bacteriophage  ($800$ $mg/ml$ or more)~\cite{Kellenberger1986}, meaning that the loops are overlapping but there is considerable space between DNA strands~\cite{Shapiro1995}. 

Previous findings strongly suggest that the loops in the network are linked once with their neighbours, and that the valence of each rings, \emph{i.e.} the number of neighbours, is around 3. In other organisms of the same class, \emph{i.e.} \emph{L. tarentolae}, the valence number is smaller, probably due to their different DNA concentration. During replication, catenation between the loops introduces a nontrivial topological problem, which is solved as follows.
First topoisomerase II disentangles one loop at a time from the network, the loop then undergoes duplication in a complex nearby, and later on it links again to the periphery of the network, together with the progeny mini-circles~\cite{Perez-Morga1993,Liu2005}. At this stage, each circle has a valence which is higher than 3: again, most likely due to the increase in density following DNA synthesis~\cite{Chen1995a}. Finally, when the cell divides, two copies of the network are produced and the valence number is brought back to 3. This change in network valence has to be mediated by topological enzymes, \emph{e.g.} topoisomerases, also accompanied by a relaxation of mass (mini-circle) density. 

The topology of the Kinetoplast DNA network is unique in its own kind and has been studied in the past with experiments and simplified models~\cite{Arsuaga2007,Diao2012}, but a full understanding of its role, origin and replication continues to represent a challenge for the scientific community~\cite{Silver1986,Morris2001,Lukes2002}.

\begin{figure*}
\centering
\includegraphics[scale=0.44]{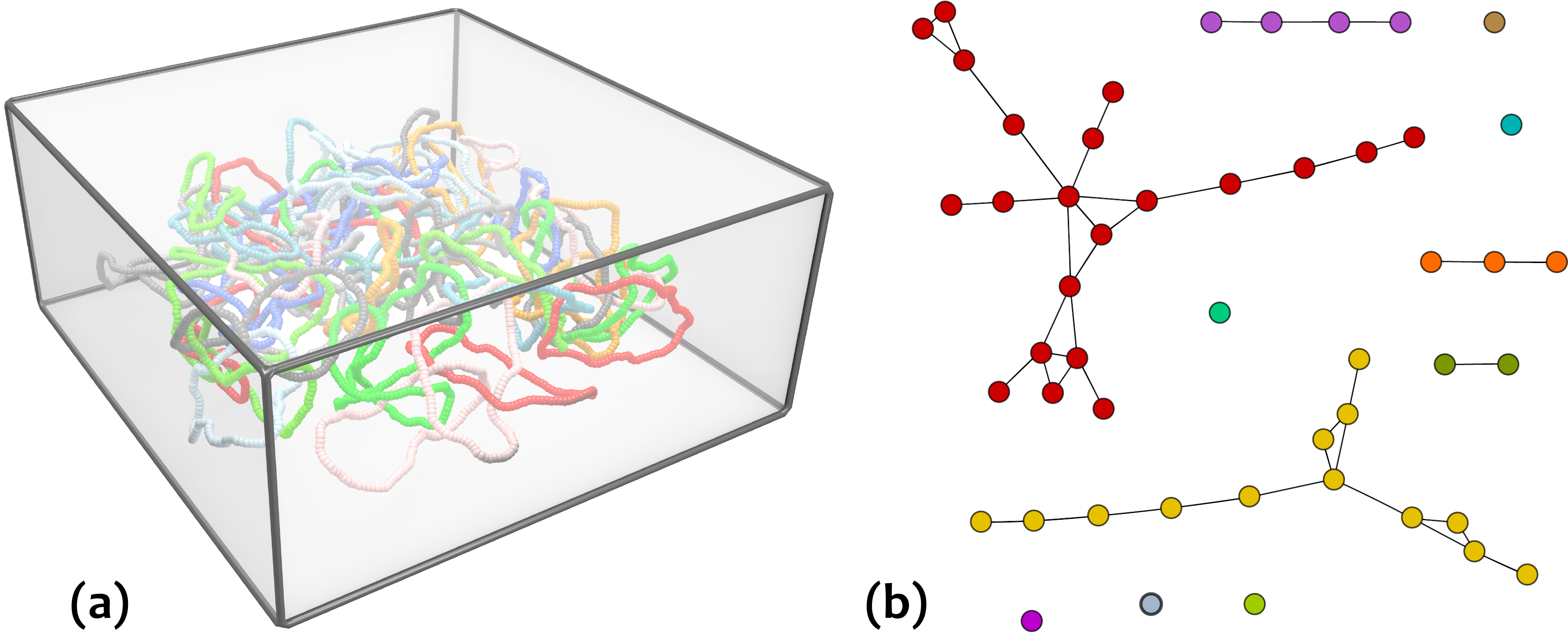}
\caption{(a) Graphical representation of the system. The colors of the rings are chosen randomly for clarity of visualisation. (b) Network representation of (a). An edge between two nodes is drawn if two rings in the system are topologically linked. Connected components are highlighted in different colors. (see text for details)}
\label{fig:panel1}
\end{figure*}

Here we propose a model which builds on fewer assumptions with respect to previous work in the literature. Here, phantom semi-flexible rings are confined to move inside a box of linear size $L$, which we vary in order to simulate varying values of the density of the Kinetoplast network inside the mitochondrial membrane. By computing the Gauss linking number between pairs of rings, we analyse the topological constraints experienced by the rings within the system. The model Kinetoplast DNA can be naturally represented as a network, by mapping rings to nodes and links between two rings to (undirected) edges~\cite{Michieletto2014,Chen1995} (see Fig.~\ref{fig:panel1}). Our findings suggest that for densities $\rho$ greater than a critical density $\rho_p$, the system has a non-zero probability of forming a cluster of linked rings as big as the size of the whole system, \emph{i.e.} percolating. In the case of the Kinetoplast, a network with this property can be viewed as a state in which a relevant fraction of the mini-circles in the Kinetoplast are mutually inter-locked to form an extended collection of inseparable rings~\cite{Diao2012}.  
We also further study the topology of the network by simulating its digestion, which is realised experimentally, for instance, by adding nuclease or restriction enzymes that cut the DNA, to the solution containing the mini-circles. Remarkably, the simulated digestion provides results which are quantitatively comparable with experiments~\cite{Chen1995}, and give new insight on the origin of the network. 

\section{Model}

We model the Kinetoplast genome as $N=50$ DNA rings, each of which is a worm-like polymer made of $M=128$ beads of size $\sigma$ and with persistence length $l_p = 20$ $\sigma$. A cut-and-shift form of the Lennard-Jones potential is used to model steric interaction between beads belonging to the same chain, so that we ensure that the rings do not get knotted and that they assume self-avoiding configurations. In physical units, $\sigma \simeq 2.5$ $nm$ is the hydrated diameter of double-stranded DNA, $l_p \simeq 50$ $nm$, while the contour length of each of the loops is $L_c = 128$ $\sigma$ $= 320$ $nm \simeq 1$ $kbp$. The network is enclosed in a box of size $L_1 \times L_1 \times L_2$, with $L_i$ between $200$ and $80$ (in units of $\sigma$). The boxes considered are both symmetric, \emph{i.e.} $L_1 = L_2$, and asymmetric, \emph{i.e.} $L_1 = 2 L_2$, such as the aspect ratio is similar to that of a Kinetoplast disk, whose thickness \emph{in vivo} ($0.4$ $\mu m$) is roughly half of its diameter ($1$ $\mu m$).

We sample different network configurations by letting the rings thermalise with no steric interaction between different rings (\emph{i.e.} rings are invisible to each other during equilibration), for at least the time taken for a ring to diffuse its own gyration radius, \emph{i.e.} $\tau_R = R^2_g/D_{CM}$. This stage mimics the presence of topological enzymes such as topoisomerases, which can either link or un-link the mini-circles from the neighbours. This allows the mini-circles to freely diffuse while temporarily unlinked. After this interval, we turn a soft repulsion on, which acts on every pair of beads distant $r$ to each other and which we model as:
\begin{equation}
E_s(t) = A(0,50;10^5)\left(1+\cos\left(\dfrac{\pi r}{r_c}\right)\right) \notag
\end{equation}
with $A(x,y;t)$ a ramp function which brings $A(x,y;t)$ from $x$ to $y$ in $t$ timesteps and $r_c = 2^{1/6} \sigma$. After this operation, which ensures that no contours are overlapping, we compute the pairwise Gauss linking number of any two rings, defined as:
\begin{equation}
Lk(i,j) = \dfrac{1}{4\pi} \int_{\gamma_i}\int_{\gamma_j} \dfrac{|\bm{r}_i - \bm{r}_j|}{|\bm{r}_i - \bm{r}_j|^3} \cdot \left(d\bm{r}_i \times d\bm{r}_j\right)
\end{equation}
where $\gamma_i$ and $\gamma_j$ are the contours of the two rings, and $\bm{r}_i$ and  $\bm{r}_j$ the respective spatial coordinates~\cite{Orlandini2000,Orlandini2004}. This is a topological invariant which describes the pairwise state of rings, as far as we forbid two rings to pass through each other. This stage represents the final Kinetoplast conformation, when no molecule of topoisomerase is available. Once the linking number has been measured, we turn the soft potential off and repeat the procedure in order to obtain an ensemble of (independent) configurations for a given density $\rho$. 

It is important to bear in mind that this ensemble of networks should be interpreted as a collection of independent equilibrated networks, rather than as a dynamical sequence of network conformations over time, since in reality the mini-circles are not allowed to cross through each other without the intervention of a topological enzyme. In other words we are generating ensembles of networks that one could obtain, for instance, when looking at the Kinetoplast after replication and after it has been separated into the two daughter cells. At this stage in fact, a simultaneous topological and structural re-arrangement of the network has to take place, involving both, topological enzymes and mass relaxation via mini-circles diffusion, which is itself allowed by the presence of topoisomerases.

The model we propose here is,  with respect to previous work, based on less assumptions, as, for example, it does not rely on the fact that the mini-circles are anchored on a 2D lattice~\cite{Diao2012}. This is an important assumption that we relax.  In fact, this rigid structure would severely compromise the Kinetoplast replication, hindering the free removal of mini-circles. We will here show that the 2D layer structure of the Kinetoplast, which is widely reported in the literature, is not required to obtain agreement with experimental observations. This suggests that condensation and anchoring into a layered 2D structure is secondary to the network topological arrangement. Furthermore, we here relax the assumption that the rings are perfect circles~\cite{Arsuaga2007}, and consider much more realistic semi-flexible polymers. 

The observables we measure from these networks are averaged over the ensemble formed by $5000$ configurations generated with the method described above. For each configuration in the ensemble, we generate a corresponding network representation by assigning an (undirected) edge between each two rings which have $Lk(i,j) \neq 0$~\cite{Michieletto2014,Chen1995}. This maps the system of linked rings to an undirected network, whose properties are directly related to the properties of the system of linked rings (see Fig.~\ref{fig:panel1}). Note that, because this procedure is based on the pairwise linking number, it would classify Borromean and Brunnian links as unlinked; we expect such non-trivial links to be rare within the Kinetoplast network, where a good approximation is that each mini-circle is linked identically and once to its neighbours~\cite{Chen1995,Jensen2012}.

Our main control parameter is the size of the confining container, $L_i$, which we modify to vary the density $\rho$. This determines the physical properties of the resulting network. The overlapping (number) density $\rho^*$, at which rings start to feel each other, can be estimated as $\rho^* = M/(4/3 \pi R^3_g) \sim  0.0076$ $\sigma^{-3}$, where $R_g$ has been measured from relaxed rings in sparse solution~\footnote{We find $R_g \sim 16$ $\sigma$, close to but below the estimate $R_g\sim\sqrt{L_cl_p/6}\sim 20.7 \sigma$ which works for $L_c\gg l_p$ and disregards excluded volume interactions within one ring.}. To convert to a biologically realistic value, we may assume that the volume occupied by each bead is that of a cylinder of size and height equal to $\sigma$, which leads to a volume fraction $\phi^* \sim 0.60 \% $ occupied by the DNA, or equivalently a concentration $c ^* \sim 8.1$ $mg/ml$ (calculated with a DNA density $\rho_{DNA} = 1.35$ $g/cm^3$~\cite{Matthews1968}).\\

\section{Results}
A good way of studying the properties of a network $G$ is by looking at its first Betti number $b_1(G)$ and its giant connected component $GCC(G)$~\cite{Michieletto2014}. The former  is defined as $b_1(G) \equiv N_{CC} - |\mathcal{V}| + |\mathcal{E}|$, where $N_{CC}$ is the number of connected components and $|\mathcal{V}|$ and $|\mathcal{E}|$ the size of the sets of vertices and edges, respectively. The latter is defined as the largest set of nodes in which every node can be reached by any other node within the set. For instance, in Fig.~\ref{fig:panel1}, the giant connected component corresponds to the red cluster. This quantity is useful to investigate the ``percolation'' of the network. Here we define a network to be percolating if the size of the giant connected component $|GCC|$ is of the same order as the number of nodes in the whole network. The percolation density $\rho_p$ is then  the density above which the system shows a non-zero probability of percolation. The giant connected component of a percolating network is a spanning, or percolating, cluster. While the size of the $GCC$ gives some information regarding the connectivity of the network, the first Betti number, $b_1(G)$, provides us with some insight about the topology of the network. In fact, $b_1(G)$ equals the number of closed sub-graphs in the network, which is also the total number of cyclic d-mers~\cite{Michieletto2014}. For mostly unconnected graphs, $b_1(G) \simeq 0$, while for nearly fully connected graphs: $|\mathcal{E}| \simeq N(N-1)/2$ and hence $b_1(G) \simeq N^2/2$ for large $N$. An increase in $b_1(G)$ corresponds to both, an increase in network connectivity and an increase in cyclic structures.

\begin{figure}[t]
\centering
\includegraphics[scale=0.6]{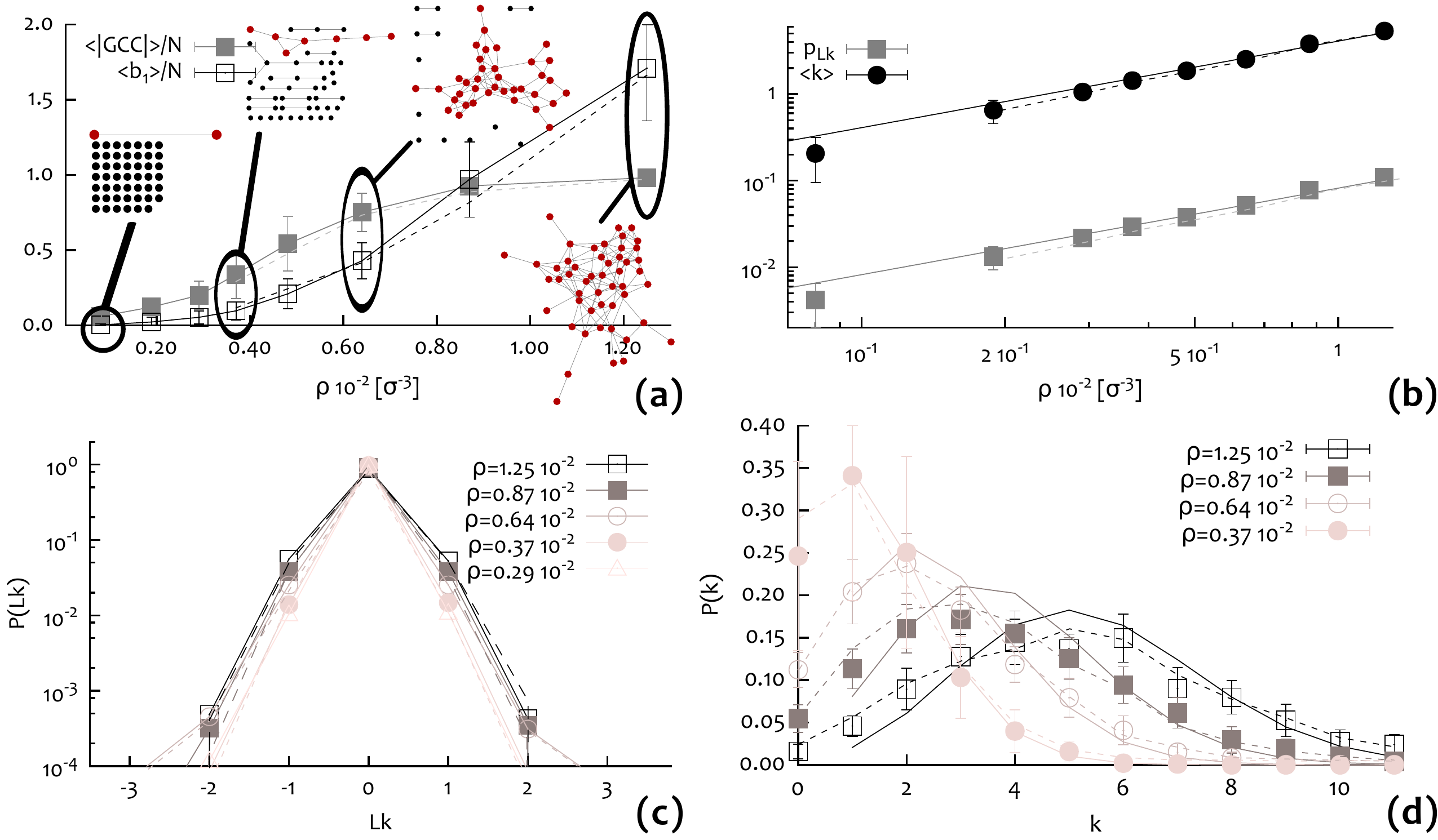}
\vspace*{-0.4 cm}
\caption{\textbf{(a)} Fraction of nodes belonging to the giant connected component $\langle | GCC | \rangle$ (filled squares) and first Betti number $\langle b_1\rangle$ (empty squares) as a function of the density $\rho$ and averaged over the ensemble of configurations. Snapshots of the typical configurations are also shown, with the respective giant connected components highlighted in red. \textbf{(b)} Valence, or mean degree, of the rings $\langle k \rangle$ together with the linking probability computed in Eq.~\eqref{eq:plk} as a function of the density $\rho$. The plot suggest a linear increase in $\rho$ in agreement with previous findings~\cite{Diao2012}. \textbf{(c)} Distribution $p(Lk)$ for different values of the density $\rho$. As found experimentally, we observe that linking numbers $Lk$ higher than $1$ and lower than $-1$ are highly suppressed. \textbf{(d)} Degree distribution $p(k)$. Data points are obtained from simulations, solid lines show the degree distributions sampled from random graphs with edge probability $p=p_{Lk}$. \textbf{(a)-(d)} Dashed lines show the results obtained from simulations performed in asymmetric boxes. (see text for details)}
\label{fig:panel2}
\end{figure}

In Fig. \ref{fig:panel2}(a) we show the size of the giant connected component, $|GCC|$, and the first Betti number of the graph $G$, $b_1(G)$, divided by the size of the system $N$ as a function of the system density $\rho$. This plot suggests that a percolating component can be observed in the system at values of the density $\rho \gtrsim \rho_p \simeq 0.0064$ $\sigma^{-3}$. Although the precise value of the percolation density is not well defined for finite systems, our model allows to predict the emergence of a state in which a fraction close to unity of rings in the system is topologically interlocked in a single cluster for values of the
density above $\rho_p$.
In Fig.~\ref{fig:panel2}(b) the valence, or mean vertex degree $\langle k \rangle$ is shown together with the linking probability $p_{Lk}$ as a function of the density $\rho$. The figure shows that the average degree $\langle k \rangle$ scales linearly with the density $\rho$, in agreement with previous findings~\cite{Diao2012}. One can also see that at $\rho=\rho_p$ the valence of our network is $3$, in agreement with experiments~\cite{Chen1995}. By assuming that the network configurations are sampled from an ensemble of random graphs, the linking probability $p_{Lk}$ can be calculated as
\begin{equation}
p_{Lk} = \langle k \rangle / (N-1).
\label{eq:plk}
\end{equation}
The assumption that the network configurations can be described by random graphs is not justified \emph{a priori}. On the the other hand, we can check that this assumption is valid by comparing the degree distribution $p(k)$ obtained from the simulations (see data points and dashed lines in Fig. \ref{fig:panel2}(d)) and the random graph distributions $p_{rg}(k;p)$ with $p = p_{Lk}$ obtained from Eq.~\ref{eq:plk}:
\begin{equation}
p_{rg}(k;p_{Lk}) = {N-1 \choose k} p_{Lk}^k (1-p_{Lk})^{N-1-k}. \notag
\end{equation}
One can notice that the random graphs distributions and the data-points are in very good agreement for $\rho \lesssim 0.0064 \sigma^{-3}$ and in agreement, but not as good, for $\rho \gtrsim 0.0087 \sigma^{-3}$; in other words, the system can be approximated as a random graph with linking probability $p_{Lk}$ which is directly proportional to the valence of the rings and inversely proportional to the number of rings in the box. In terms of the Kinetoplast structure one can imagine that, due to the presence of topological enzymes, the mini-circles can be un-linked and hence undergo free diffusion inside the mitochondrion. Because of this, it is reasonable to expect that the network of mini-circles would form a random arrangement of linked rings. In this respect, our model can capture the randomness of the system in a better way than previous models found in the literature could.

In Fig. \ref{fig:panel2}(c) we show the distribution of the linking number $p(Lk)$. This quantity is found to be peaked at zero for any density $\rho$ investigated in this work. This means that the networks produced by our model are never fully connected, \emph{i.e.} there are always more pairs of rings which are unlinked than pairs which are linked. The ``shoulders'' of the distribution at $Lk = \pm 1$ increase with $\rho$ although values of $|Lk|>1$ are very unlikely ($p(Lk) < 10^{-3}$). This is once again in agreement with experimental findings~\cite{Chen1995,Jensen2012}, which observed that each linked mini-circle is linked only once with its neighbours. 
For simplicity, we always assign a single edge between a pair of nodes even in those cases in which they have an higher linking number. Because they are so rare, they represents a small fraction of all the links, which can be neglected. The mean linking number $\langle Lk \rangle$ is zero within errors, as must be the case as configurations with $Lk=-1$ are as likely as ones with $Lk=+1$. 

It is worth noting that in Figs.~\ref{fig:panel2}(a)-(d) we  report the results obtained by simulating the system in asymmetric boxes. These are shown as dashed lines. As one can notice, the two cases are in very good qualitative agreement. Small deviations are found for the degree distributions and the average Betti number $\langle b_1 \rangle$. This suggests that the actual shape of the confining box does not affect the qualitative behaviour of the system, which preserves its randomness.

\begin{figure}[t]
\centering
\includegraphics[scale=0.65]{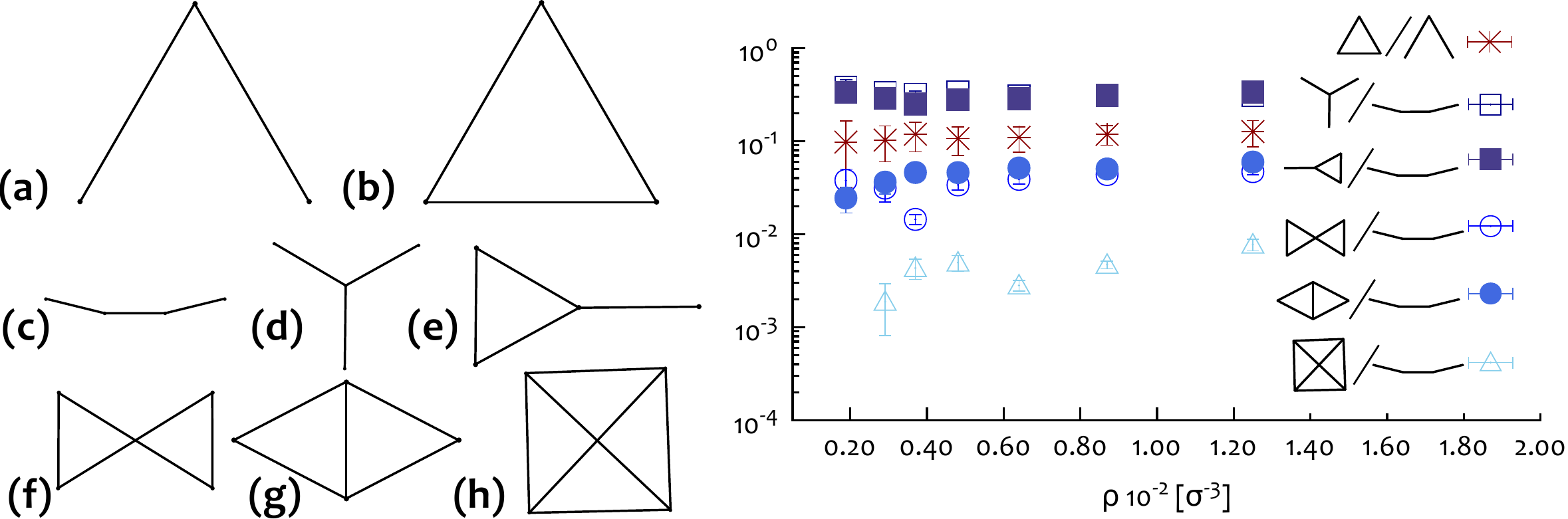}
\caption{Showing the fraction of cyclic trimers \textbf{(b)} over linear trimers \textbf{(a)} (red) and the fraction of different tetramers (\textbf{(d)}-\textbf{(h)}) over linear tetramers \textbf{(c)} (shades of blue). The findings are in qualitative agreement with previous experiments~\cite{Chen1995}.}
\label{fig:motif3_4}
\end{figure}

In Fig. \ref{fig:motif3_4}(a)-(h) we show all the possible sub-graphs with three and four nodes, up to symmetries. We call these patterns ``motifs''. Every connected graph formed by three and four nodes is isomorphic to those in Fig. \ref{fig:motif3_4}(a)-(b) and Fig. \ref{fig:panel1}(c)-(h), respectively. In order to count the number of motifs of each type, we consider every connected sub-graph with given number of nodes and check whether it is isomorphic to one of the motifs shown in Fig. \ref{fig:panel1}. The results are shown in Fig. \ref{fig:motif3_4}. We observe that linear trimers (motif \ref{fig:motif3_4}(a)) are much more frequent than cyclic ones (motif \ref{fig:motif3_4}(b)), for any density $\rho$. Similarly, linear tetramers, are three times more common than branched tetramers (motif \ref{fig:motif3_4}(d)), which is in qualitative agreement with experiments~\cite{Chen1995}. Finally, fully cyclic tetramers are highly suppressed, again in agreement with previous experimental work.

\begin{figure}[t]
\centering
\includegraphics[scale=0.62]{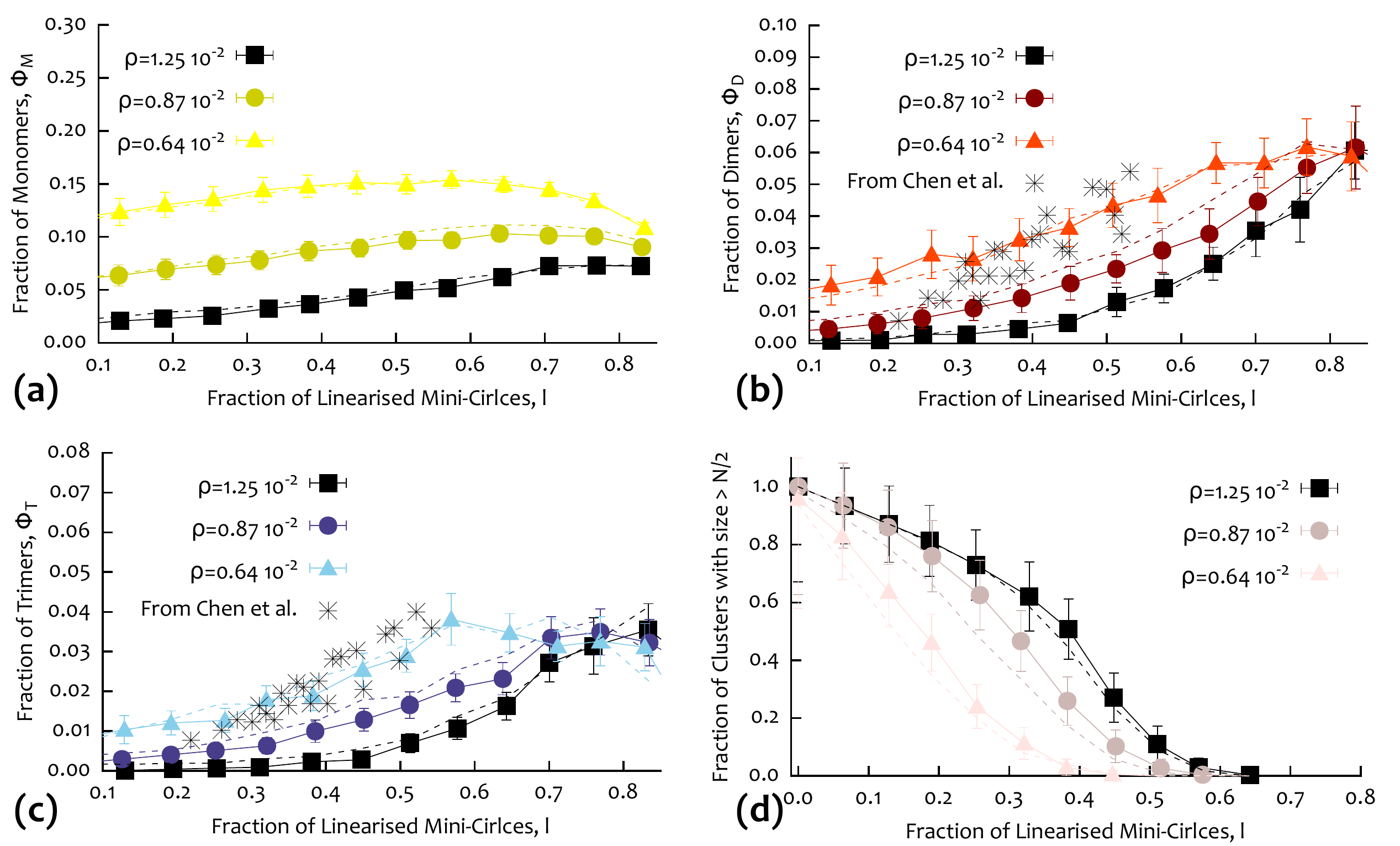}
\caption{From top to bottom: Fraction of Monomers, Dimers, Trimers and Clusters with size $s>N/2$ as a function of the fraction of mini-circles linearised during the digestion, for different densities $\rho$. Crosses represent experimental data from Ref.~\cite{Chen1995}. Dashed lines represent values obtained by performing simulations in asymmetric boxes.}
\label{fig:Digest}
\end{figure}

To further quantitatively compare the properties and structure of the random network of linked confined loops found in simulations to those of the Kinetoplast DNA, and inspired by the common biological procedure known as ``digestion'', we simulate the presence in solution of restriction enzymes, \emph{i.e.} enzymes which are able to cut DNA strands. 
In this way we can simulate the random breakage of the network due to a concentration of restriction enzymes which can cut the network, and which was used in Ref.~\cite{Chen1995} to further study the network topology experimentally. 
To model digestion {\it in silico}, we associate a probability $p$ to each bead composing the rings to be removed. Such probability is related to the concentration of restriction enzymes in the solution and time left to act on DNA mini-circles. The equivalent probability $p_r$ of a ring to become linearised, \emph{e.g.} by removing one or more of its beads, is
\begin{equation}
p_r = 1 - (1-p)^M \sim M p
\end{equation} 
where $M$ is the number of beads composing the rings. For probability $p = p_c = 1/M$, every ring has been cut, on average, once, and therefore there are no longer closed rings in the system. This procedure maps to the network representation as we can assign the same probability $p_r$ to each of the nodes, and with probability $p_r$ we remove a node from the network. In practice, we consider an ensemble of (independent) configurations from the Molecular Dynamics simulation and for each one we simulate 10 digestions by removing nodes at random with probability $p_r$ from the corresponding network. We then average the observables over the ensemble of $50 \times 10^3$ simulated digestions. The average number of removed nodes is $\langle n_r \rangle = p_r N$. At the end of the (partial) digestions we measure the fraction of monomers (single uncatenated rings), dimers (two catenated rings) and trimers (three catenated rings) obtained from the digested network. These quantities can be obtained by running a high resolution gel electrophoresis test on the samples, as in \cite{Chen1995}. The relative fraction of monomers, dimers, trimers etc, correlates directly with the intensity of the bands, as these oligomers move with different speed in the gel. In Fig. \ref{fig:Digest} we report our findings for different values of density $\rho$, as a function of the linearised fraction of mini-circles, $l = \langle n_r \rangle$ (to ease comparison with the data in Ref.~\cite{Chen1995}).

From the results in Fig.~\ref{fig:Digest}(d), one can notice that even after half of the nodes have been removed, one can still observe some large, or percolating, clusters. In other words, the network shows high resistance against random breakage. When viewed as a property of the biological Kinetoplast genome, this appears to be functionally relevant: the DNA network needs to remain intact either when some of the mini-circles are removed, e.g. by topoisomerase II, either accidentally during the cell cycle, or during replication when decatenation is required.

The distribution of the fraction of monomers $\Phi_M$, dimers $\Phi_D$ and 
trimers $\Phi_T$ show peaks as a function of $l$, whose locations depend on the 
density. In general, the value of $\rho$ at which the distributions reach their 
maximum increases with density, meaning that the denser the system, the more it 
has to be digested before the probability of observing monomers, dimers or 
trimers, rises and becomes sizeable. For fixed density, the peaks show that 
trimers are best produced at lower $l$ than dimers, and dimers at lower $l$ than 
monomers; this is expected as increasing $l$ should increase the probability of 
finding smaller and smaller catenanes. We find that the best range of $l$ within 
which gel electrophoresis of oligomers can give information on the network 
structure depends on the density $\rho$. For the highest density studied here, 
the fraction of linearised mini-circles has to be close to $80\%$, while for 
$\rho \sim 0.0064$ $\sigma^{-3}$ the value of $l$ can lay between $30\%$ and 
$80\%$, after which the survival fraction of dimers and trimers start to 
decrease. This range is very similar to the one observed in 
Ref.~\cite{Chen1995}. Even more strikingly, in Fig.~\ref{fig:Digest} we 
superimpose the data from Ref.~\cite{Chen1995}, and observe a striking 
quantitative agreement with the curve for $\rho = 0.0064$ $\sigma^{-3}$ -- we 
recall that $\rho = 0.0064$ $\sigma^{-3}$ also leads to $\langle k \rangle 
\simeq 3$ as inferred from the experiments.  Remarkably, considering the 
asymmetric system results in very little difference with the curves reported in 
Fig.~\ref{fig:Digest}. This strongly suggests that the simple symmetric 
confinement is enough to understand the Kinetoplast structure both qualitatively 
and quantitatively. 


The good agreement with experimental data shown in Fig.~\ref{fig:Digest}, strongly 
suggest that our model  can capture the topological structure of the Kinetoplast DNA described by a network of randomly connected confined 3D links close to the percolation 
transition. In this respect, the fact that the Kinetoplast DNA is found geometrically to be a layered disk-shaped structure may be due to a combination of the geometrical spatial confinement the network is subject to {\it in 
vivo}~\cite{Lukes2002,Ogbadoyi2003,Gluenz2007a,Lai2008,Docampo2010,Diao2012} and of the action of histone-like DNA-binding 
proteins~\cite{Xu1993,Xu1996,Hines1998,Avliyakulov2004}. However, the latter is inessential to 
explain the existing digestion data. 
In addition, in the Supplementary Material we show that this layered organised structure can be achieved within our framework by adding suitable interactions between some parts of the rings and the confining box (see S.I. for details).

One of the main conclusions that can be drawn from our work is that the Kinetoplast topology is independent on the network packaging and organisation, while it is driven solely by geometrical confinement. This prediction can be tested, for instance, by measuring the valence of \emph{in vivo} networks, as done in Refs.~\cite{Chen1995,Chen1995a}, formed when genes expressing KAP histone-like proteins are silenced, as done in Ref.~\cite{Avliyakulov2004}. Our results predict that in this case, the Kinetoplast should appear un-layered and disorganised, while retaining a valence near 3.

\section{Discussion and conclusions}

In summary, we studied the statistical physics of a percolating cluster of linked rings, by confining phantom semiflexible rings in a box and varying the density. The onset of the percolation occurs at concentrations $\rho_p \sim 0.0064$ $\sigma^{-3}$.
At this density, the mean valence of the nodes is around $3$, which is compatible with the findings in the Kinetoplast DNA. Importantly, at the same density value, we compared the results from an {\it in silico} digestion of the network by a restriction enzyme, finding very good quantitative agreement with the experimental data found in Ref.~\cite{Chen1995}. These results strongly suggest that the Kinetoplast topology is well represented by this model at density $\rho \sim \rho_p$, i.e. by a network of linked rings close to its critical point, \emph{i.e.} the point at which the network starts to show percolating behaviour. Remarkably, our results are affected very little by the details of the confining geometry -- what matters is the presence of confinement itself, which drives the percolation transition in the network of links~\cite{Diao2012}. 

Our findings also suggest that the density of DNA loops in the Kinetoplast networks should not be too far from the overlap density. Taking a typical case with $N=5000$ loops of say $1$ $kbp$ each, we find that the overlap density is $\sim 8.1$ $mg/ml$; the density of the same network within a mitochondrion of volume $\sim 1$ $\mu m^3$ volume is about $5.43$ $mg/ml$, which fits very well with our simulations. The DNA structure in \emph{C. fasciculata}, which is well studied, has larger density ($\sim 50$ $mg/ml$), but this is achieved by further compaction by histone-like proteins~\cite{Hines1998,Avliyakulov2004}, hence does not reflect purely geometric confinement. Furthermore, even if the density is larger than the overlap density, a network could still exist close to the percolation transition if the activity of topoisomerase II, which allows catenation and is tacitly assumed by our model as loops are invisible to each other, is limited, for instance by the enzymatic concentration. 

Being close to the percolation transition may well provide an evolutionary advantage for the Kinetoplast DNA network, as this structure may be favoured over a more heavily connected network, as it facilitates the decatenation during replication, but at the same time ensures that mini-cirlces are not released by mistake. Another property of the Kinetoplast-like network is that it is very resistant to digestion by a restriction enzyme, \emph{i.e.} the digestion has to proceed significantly before large clusters disappear (see Fig.~\ref{fig:Digest}). This feature again appears to be functionally relevant, as it provides a way to preserve genetic material against random breakage and replication mistakes.

\section{ACKNOWLEDGEMENTS}
DMi acknowledges the support from the Complexity Science Doctoral Training Centre at the University of Warwick with funding provided by the EPSRC (EP/E501311). EO acknowledge financial support from the Italian ministry of education grant PRIN 2010HXAW77. We also acknowledge the support of EPSRC to DMa, EP/I034661/1. The computing facilities were provided by the Centre for Scientific Computing of the University of Warwick with support from the Science Research Investment Fund.


\appendix 
\section{Supplementary Information}

\begin{figure}[h!]
\centering
\includegraphics[width=1\textwidth]{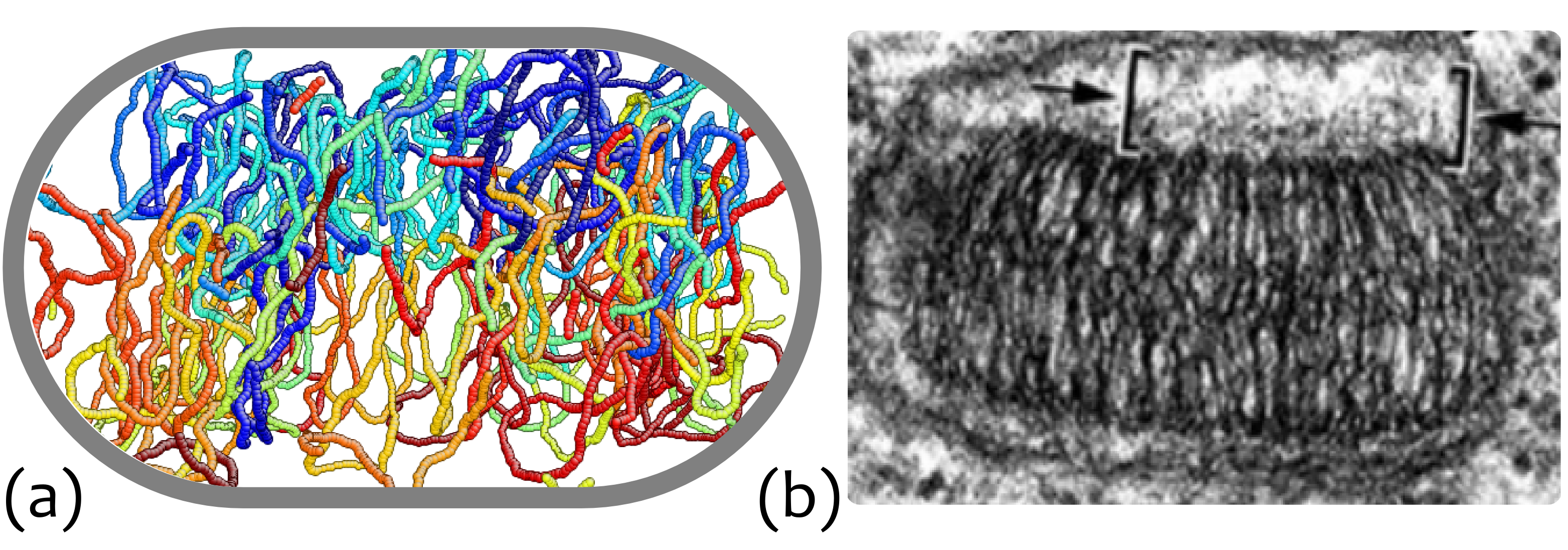}
\caption{(a) Cross-section image from MD simulations where part of the rings are attracted either to the top (shades of blue) or bottom (shades of red) surface of the confining envelope. (b) Microscopy cross-section image from \emph{C. fasciculata} (Ref.~\cite{Avliyakulov2004}).}
\label{fig:Layer}
\end{figure}

The Kinetoplast DNA is often observed to form a layered structure, where mini-circles are positioned in one or two tiers (see Fig.~\ref{fig:Layer}(b) which shows a microscopy image from \emph{C. fasciculata}). Such ordered packaging has often led the scientific community to interpret the Kinetoplast as a two-dimensional network of rings. In Fig.~\ref{fig:Layer} we show that such structure can be obtained from a 3D arrangement of rings (obtained as described in the main text) by inducing an attraction between the rings and the surrounding environment. Here we show cross-sectional snapshots from MD simulations performed in asymmetric boxes where every ring in the confining geometry possesses one bead having an attraction either to the top (shades of blue) or bottom (shades of red) surface forming the confining structure. This mimics the presence of anchoring proteins with specific bindings to, for instance, bent sequences~\cite{Silver1986}. As one can observe, by simply adding this attraction after the topological re-arrangement of the network, our model can achieve the layered structure that is usually observed in the \emph{in vivo} Kinetoplast DNA (see Fig.~\ref{fig:Layer}).

We argue that this qualitatively supports the fact that such packaging can occur after the topological organisation of the Kinetoplast, and that our model captures all the essential features explaining the structure of the Kinetoplast. 


\bibliographystyle{iopart-num}
\bibliography{RandomLinking_mod}

\end{document}